\documentclass{article}
\usepackage{fullpage,amsthm,amsfonts,amsmath}

\newtheorem{theorem}{Theorem}
\newtheorem{lemma}{Lemma}
\newtheorem{corollary}{Corollary}

\newcommand{\ket}[1]{\lvert #1 \rangle}

\newcommand{\C}{\mathbb{C}}
\DeclareMathOperator{\supp}{\mathsf{supp}}

\newcommand{\suppress}[1]{}
\newcommand{\journal}[1]{}

\def\Z{{\mathbb Z}}
\def\F{{\mathbb F}}

\newcommand{\HSP}{\mbox{\textsc{HSP}}}
\newtheorem{fact}{Fact}

\DeclareMathOperator*{\central}{\mathbf{Y}}
\newcommand{\multicentral}[3]{\central_{#1}^{#2} #3}
\newcommand{\prob}{\mathop{\mathsf{Pr}}}
\begin{document}

\title{An efficient quantum algorithm for  the hidden subgroup problem
in extraspecial groups
\thanks{
Research supported by the European Commission IST Integrated Project
Qubit Applications (QAP) 015848, the OTKA grants T42559 and T46234, 
the NWO visitor's grant Algebraic Aspects of Quantum Computing,
and by the 
ANR Blanc AlgoQP grant of the French Research Ministry. 
%\vfil
}}

\author{
G\'abor Ivanyos\footnotemark[2]
\and
Luc Sanselme\footnotemark[3]
\and
Miklos Santha\footnotemark[4]
}

\maketitle

\renewcommand{\thefootnote}{\fnsymbol{footnote}}

\footnotetext[2]{SZTAKI, Hungarian Academy of Sciences,
H-1111 Budapest, Hungary.
{\tt ivanyos}{\tt @sztaki.hu}}

\footnotetext[3]{UMR 8623 Universit\'e Paris--Sud
91405 Orsay, France.
{\tt sanselme@lri.fr}
}

\footnotetext[4]{CNRS--LRI, UMR 8623 Universit\'e Paris--Sud
91405 Orsay, France.
{\tt santha@lri.fr}
}

\begin{abstract}
%Gabor: made the abstract somewhat more modest
Extraspecial groups form a remarkable subclass of $p$-groups.
They are also present in quantum information theory, in particular 
in quantum error correction. We give here a polynomial time
quantum algorithm for finding hidden subgroups in extraspecial groups. 
Our approach is quite different from the recent
algorithms presented in \cite{rrs05} and \cite{bcv05}
for the Heisenberg group, the extraspecial $p$-group of size
$p^3$ and exponent $p$. Exploiting certain nice automorphisms of the 
extraspecial groups we define specific
group actions which are used to reduce the problem to hidden subgroup instances 
in abelian groups that can be dealt with directly.
\end{abstract}
%{\bf Key words:} Quantum computing, efficient algorithm, group theory

\section{Introduction}
The most important challenge of quantum computing is to find quantum algorithms
that achieve exponential speedup over the best known classical solutions.
In this respect, the most extensively studied problem 
is the paradigmatic hidden subgroup problem.
Stated in a group theoretical setting,
in $\HSP(G,f)$
we are given explicitely a finite group $G$ and 
we also have at our disposal 
a function $f$ that can be queried via an oracle, and which
maps  $G$ into a finite set.
We are promised that for some subgroup $H$,
$f$ is constant on each left coset of $H$
and distinct on different left cosets.
We say that $f$ {hides} the subgroup $H$.
The task is to determine the {hidden subgroup} $H$.
We measure the time complexity of an algorithm  
by the overall running time when a query counts as one
computational step.
An algorithm is called {efficient}
if its time complexity
is polynomial in the logarithm of the order of $G$.

We don't know any classical algorithm of polynomial query complexity 
for the HSP, even in
the restricted case of abelian groups.
In this respect, probably the most important result of 
quantum computing is that the HSP can be solved 
efficiently for abelian groups by quantum algorithms.
We will call this solution, for which
one can find
an excellent description for example
in Mosca's thesis~\cite{mos99}, the standard algorithm
for \HSP{}.
The main quantum tool used in the standard algorithm 
is Fourier sampling based on
the approximate quantum Fourier transform that can be efficiently
implemented by a quantum algorithm in case of abelian
groups~\cite{kit95}.
Among the important special cases of this general solution one can mention
Simon's xor-mask finding~\cite{sim97},
Shor's factorization
and discrete logarithm finding algorithms~\cite{sho97},
and Kitaev's algorithm~\cite{kit95} for the abelian stabilizer problem.

Since the realization of the importance of the abelian HSP,
intensive efforts have been made to solve the hidden subgroup problem also
in finite non-abelian groups. The intrinsic mathematical interest of this challenge is 
increased by the fact that several famous classical algorithmic problems can be cast
in this framework, like for example 
the graph isomorphism problem.
The successful efforts for solving the problem can roughly be divided into two categories. 
The standard 
algorithm has been extended to some non-abelian groups
by R\"otteler and Beth~\cite{rb98}, Hallgren, Russell and Ta-Shma~\cite{hrt03},
Grigni, Schulman, Vazirani and Vazirani~\cite{gsvv01} and 
Moore, Rockmore, Russell and Schulman~\cite{mrrs04}
%in \cite{rb98,hrt00,gsvv01,mrrs04} 
using efficient implementations
of the quantum Fourier transform
over these groups. 
%Unfortunately, efficient quantum Fourier transform
%implementations are known only for a few non-abelian 
%groups~\cite{bea97,prb99,rb98,hrt00}.
In a different approach, Ivanyos, Magniez and Santha~\cite{ims03} and
Friedl, Ivanyos, Magniez, Santha and Sen~\cite{fimss03} have efficiently reduced
the  \HSP{} in some non-abelian groups to \HSP{} instances in abelian groups
using classical and quantum group theoretical tools, but not the 
non-abelian Fourier transform.
%was efficiently solved in the context of specific non-abelian black-box groups~\cite{bs84,wat01}
%by~\cite{ims03} without using the Fourier transform on the group.

All groups where the HSP has been efficiently solved are in some sense ``close"
to abelian groups. Extraspecial groups, in which we present here an efficient 
quantum algorithm,
%for the HSP, 
are no exception in this respect: they have the property
that all their 
%Gabor: inserted
proper 
factor groups are abelian. They form a subclass of
$p$-groups, where $p$ is a prime number,
and play an important role in the theory of this family of groups. 
Extensive treatment
of extraspecial groups can be found for example in the books of Huppert~\cite{hup83} and
Aschbacher~\cite{asc00}.

Extraspecial 2-groups are heavily present in the theory of quantum error correction.
They provide a bridge between quantum error correcting codes  and binary 
orthogonal geometry~\cite{crss97}. They form the real subgroup of the Pauli group~\cite{crss98}
which plays a crucial role in the theory of stabilizer codes~\cite{got97}.
%Gabor:
For general $p$, extraspecial $p$-groups give rise to
the simplest examples of Clifford codes, see~\cite{ks06}.

Efficient solutions for the HSP have already been given in several specific extraspecial groups.
Extraspecial $p$-groups are of order $p^{2k+1}$ for some integer $k$.
For odd $p$, they are of exponent $p$ or $p^2$, and
extraspecial 2-groups are of exponent 4.
The class of groups for which Ivanyos, Magniez and Santha~\cite{ims03} provide a solution
include extraspecial $p$-groups when $p$ is a fixed constant and the input size grows with $k$.
When $p$ is fixed, the smallest 
%Gabor: abelian groups have trivial commutators
%non-abelian  
extraspecial groups
are of size $p^3$. Up to isomorphism there are two extraspecial groups of order $p^3$.
%one of them is of exponent $p$, the other of exponent $p^2$.
Recently two independent works dealt with quantum algorithms
for the HSP in the group of exponent $p$, the Heisenberg group. 
Radhakrishnan, R\"otteler and Sen~\cite{rrs05} have followed the standard algorithm
with non-abelian Fourier transform, and proved that strong Fourier sampling with a random
basis leads to a query efficient quantum solution. In a subsequent work, 
Bacon, Childs and van~Dam~\cite{bcv05} devised an efficient quantum algorithm, where
%have used 
%abelian Fourier transform and 
a state estimation technique, called the pretty good
measurement, is used to reduce the HSP to some matrix sum problem that they could solve
classically.

In this paper we provide an efficient quantum algorithm for the HSP in any extraspecial group.
Our main contribution is an efficient algorithm in extraspecial $p$-groups of exponent $p$
when $p$ grows with the input size. A simplified version of this algorithm gives another solution
for the groups of constant exponent. The remaining case, groups of exponent $p^2$ when $p$ is
large is easily reducible to the case of groups of exponent $p$.

Our approach for groups of exponent $p$ is completely different 
from the above two solutions for the Heisenberg group.
In our solution only abelian Fourier transforms and von Neumann measurements are used.
In fact, our algorithm is a series of reductions, where we repeatedly use 
%an extension of 
the standard
algorithm for abelian groups, or a slight extension of it.
In this extension, instead of a classical hiding functions we have an
efficient quantum hiding procedure at our disposal. 
This procedure outputs a quantum state for every group
element so that the states corresponding to group elements coming from the same left coset of
the hidden subgroup are identical, whereas the states corresponding to group elements from different
left cosets are orthogonal. Repeated invocations of the procedure might yield different states for
the same group element.

At the end of our reductions we are faced with the problem of creating an efficient hiding
procedure in the above sense for the subgroup $HG'$ of $G$, where $G$ is an extraspecial 
$p$-group of exponent $p$ when $p$ is large,
$G' = \{z^i ~:~ 0 \leq i \leq p-1 \}$
is its commutator, and $H$ is the hidden subgroup. It is easy to see, that if we could create
the coset state $\ket{aHG' }$ for some $a \in G$, then the group action multiplication from the right,
which on a given group element 
$g$ would output $\ket{aHG' \cdot  g}$, is a hiding procedure. 
Unfortunately, we can create these states efficiently 
only when $p$ is constant. In the general case, we can create efficiently only the states 
$ \ket{aHG'_u}$ for a random $ 0 \leq u \leq p-1$, where 
$\ket{G'_u}=
\frac{1}{\sqrt{p}} \sum_{i \in \Z_p} \omega^{-u i} \ket{z^i}$.
Our main technical contribution is to show that several (in fact four) copies of these states can be 
combined together so that the disturbing phases cancel each other. To achieve this goal
we exploit 
certain nice automorphisms
%the automorphism structure 
of the group to define more sophisticated group actions
that can be used for our purposes.

The structure of the paper is quite simple. After a discussion on the extension of the standard 
algorithm and a basic description of extraspecial groups in Section~\ref{prelim}, our reduction
steps are presented in Section~\ref{reductions}. The summary of these reductions is stated in
Theorem~\ref{theorem:threelemmas}: An efficient hiding procedure for $HG'$ is sufficient
to solve the HSP in an extraspecial group $G$. In Section~\ref{algorithm} we establish our main 
result in Theorem~\ref{theorem:main}, the existence of an efficient solution for the HSP in extraspecial
groups. The proof is given according to the three cases discussed above. The most important case
of groups of exponent $p$ when $p$ is large is dealt with in Section~\ref{important}, where in 
Theorem~\ref{theorem:exponentp} we provide the hiding procedure for $HG'$.

\section{Preliminaries}\label{prelim}

\subsection{Extensions of the standard algorithm for the abelian \HSP}

We will use standard notions of quantum computing for which one can consult for 
example~\cite{nc00}.
For a finite set $X$, we denote by $\ket{X}$ the uniform superposition
$\frac{1}{\sqrt{|X|}}\sum_{x \in X} \ket{x}$ over $X$.
For a superposition $\ket{\Psi}$, we denote by $\supp(\ket{\Psi})$ the support of $\ket{\Psi}$,
that is the set of basis
elements with non-zero amplitude.

The general solution for the abelian $\HSP$ consists essentially
of Fourier sampling of the hiding function $f$. More specifically, it involves 
the creation
of the superposition $\sum_{g \in G} \ket{g}\ket{f(g)}$ and
the Fourier transform over $G$. 
Clearly, for the former part it is essential to have access to a hiding function. In fact, this
requirement can be relaxed in some sense, and in this paper we will use such a relaxation.
A relaxation was already used by Ivanyos et al.~\cite{ims03} who extended the notion of the hiding
function to quantum functions. More precisely, for a finite set $X$, and a quantum function 
$f : G \rightarrow {\C}^X$, we say that $f$ {\em hides} the subgroup $H$ of $G$ if $\ket{f(g)}$
is a unit vector for every
$g \in G$, and $f$ is constant on the left cosets of $H$,
and maps elements from different cosets into orthogonal states. The simple fact is proven
in Lemma 1 of \cite{ims03} that in the standard solution of $\HSP$ for abelian groups, one 
can just as well use a quantum hiding function.

The standard algorithms for the abelian $\HSP$ in fact repeats 
polynomially many times the Fourier sampling
involving the same (classical or quantum) hiding function. 
In fact, in each iteration a random element is obtained from the subgroup orthogonal to $H$.
Our extension is based on the observation, that for the sampling, one doesn't have to use the
same hiding function in each iteration, different hiding functions will do just as well the game.
For the sake of completeness we formalize this here and state the exact conditions that will
be used in our case.

We say that a set of vectors $\{ \ket{\Psi_g} : g \in G \}$ from some Hilbert space $\cal{H}$ is 
a {\em hiding set} for
the subgroup $H$ of $G$ if
\begin{itemize}
\item
$\ket{\Psi_g}$ is a unit vector for every $g \in G$,
\item
if $g$ and $g'$ are in the same left coset of $H$ then $\ket{\Psi_g} = \ket{\Psi_{g'}}$,
\item
if $g$ and $g'$ are in different left cosets of $H$ then $\ket{\Psi_g}$ and $\ket{\Psi_{g'}}$
are orthogonal.
\end{itemize}
A quantum procedure is {\em hiding} the subgroup $H$ of $G$ if
for every $g \in G$, on input $\ket{g}\ket{0}$ it outputs $\ket{g}\ket{\Psi_g}$ where
$\{ \ket{\Psi_g} : g \in G \}$ is a hiding set for $H$. 
Let us underline that we don't require from a quantum hiding procedure to output
the same hiding set in different calls.
The following fact  recasts
the existence of the standard algorithm for the abelian $\HSP$ in the context of hiding sets. \begin{fact}\label{fact:hsp}
Let $G$ be a finite abelian group. If there exists an efficient quantum procedure 
which hides the subgroup $H$ of $G$ then there is an efficient quantum algorithm for
finding $H$.
\end{fact}
\begin{proof}
It is immediate from the proof of Lemma 1 in \cite{ims03}: indeed, the exact property of the
quantum hiding function $f$ which is used there is that $\{\ket{g} \ket{f(g)} : g \in G \}$ forms
a hiding set for $H$.\qed
\end{proof}

\subsection{Extraspecial groups}

Let $G$ be a finite group. For two elements $g_1$ and $g_2$ of $G$, we usually denote their product
by $g_1g_2$. If we conceive group multiplication from the right as a group action of $G$ on itself, we
will use the notation $g_1 \cdot g_2$ for $g_1g_2$. For a subset $X$ of $G$, we will denote by 
$\langle X \rangle$ the subgroup generated by $X$. The derived subgroup $G'$ of $G$ is defined as
$\langle \{ x^{-1}y^{-1}xy ~:~ x,y \in G \} \rangle$, and its center $Z(G)$ as
$\{z \in G ~:~ gz = zg \mbox{{\rm ~ for all ~}} g \in G \}$.
The Frattini subgroup $\Phi(G)$ is the intersection of all maximal subgroups of $G$.

For an integer $n$, we denote by $\Z_n$ the group of integers modulo $n$,
and for a prime number $p$, we denote by $\Z_p^*$ the multiplicative group of integers 
relatively prime with $p$. A $p$-group is a finite group whose order is a power of $p$.
A $p$-group $G$ is {\em extraspecial} if $G' = Z(G) = \Phi(G)$, and its center is cyclic of prime order $p$.

If $G$ is an extraspecial $p$-group then $|G| = p^{2k+1}$ for some integer $k$. 
%Gabor: replaced
The elements of $G$ can be encoded by binary strings of length  
%The encoding length of such a group is 
$O(k \log p)$, 
and an efficient algorithm on that input
has to be polynomial in both $k$ and $\log p$. 
%We will consider the case of extraspecial
%$p$-groups where $p$ is a constant, and also the case where $p$ goes to infinity.

The smallest non-abelian extraspecial groups are of order $p^3$. For $p=2$, we have, up to
isomorphism, two extraspecial 2-groups of order 8.  These are the quaternion group $Q$, and the
dihedral group $D_4$, the symmetry group of the square in two dimensions.
The exponent of both of these groups is $p^2 = 4$.

For $p > 2$, up to isomorphism we have again two 
extraspecial $p$-groups of order $p^3$. The first one is the
Heisenberg group $H_p$, which is the group of upper triangular 
$3 \times 3$ matrices over the field $\F_p$
whose diagonal contains everywhere 1. The exponent of $H_p$ is $p$. The other one is $A_p$,
the group of applications $t \mapsto at +b$ 
from $\Z_{p^2}$ to $\Z_{p^2}$, where $a \equiv 1$ modulo $p$
and $b \in \Z_{p^2}$. The exponent of $A_p$ is $p^2$.

We give now via relations equivalent definitions of the extraspecial $p$-groups of order $p^3$.
These definitions will be useful for the arguments we will develop in our algorithms. To
emphasize the similarities between these groups, we will take three generator elements
$x,y,z$ for each of them.
%, even though in all the groups but in $H_p$ already two generators are sufficient. 
The element
$z$ will always generate the center of the group. Here are the
definitions via relations:
$$Q_{} = \langle x^2 = y^2 = [x,y] = z, ~z^2 = 1 \rangle ,$$
$$D_{4} = \langle x^2 = y^2 = z^2 =1, ~[x,y] = z, ~[x,z] = [y,z] = 1 \rangle ,$$
$$H_{p} = \langle x^p = y^p = z^p =1, ~[x,y] = z, ~[x,z] = [y,z] = 1 \rangle ,$$
$$A_{p} = \langle x^{p^2} = y^p =1, ~[x,y] = z = x^p, ~[y,z] = 1 \rangle .$$
From these definitions it is clear that every element in an extraspecial group of order $p^3$ 
has a unique representation of the form $x^iy^jz^{\ell}$ where $i,j,\ell \in \Z_p$.

Extraspecial $p$-groups of order $p^{2k+1}$, for $k > 1$, can be obtained as the central product of $k$
extraspecial $p$-groups of order
$p^3$. If $G_1, \ldots, G_k$ are extraspecial $p$-groups of order $p^3$ then their 
{\em central 
product} ~$G_1 \central \ldots \central G_k$ is the 
%Gabor:
factor
%quotient 
group
$$G_1 \times \ldots \times G_k \bmod z_1 = \dots = z_k,$$
where $z_i$ is an arbitrary generator of $Z(G_i)$ for $i = 1, \ldots, k.$

Since $D_4 \central D_4 = Q \central Q$, up to isomorphism the unique extraspecial 2-groups
of order $2^{2k+1}$ are $\multicentral{i=1}{k}{D_4}$ and 
$(\multicentral{i=1}{k-1}{D_4}) \central Q.$
All of these groups are of exponent $p^2 = 4$. 
%REMARK WE SHOULD ADD: $\multicentral{i=1}{k}{D_4}$ is 
%the group used in quantum error correcting.
When $p > 2$, we have 
$H_p \central A_p = A_p \central A_p$. Therefore, 
up to isomorphism the unique extraspecial $p$-groups
of order $p^{2k+1}$ are $\multicentral{i=1}{k}{H_p}$ and $(\multicentral{i=1}{k-1}{H_p}) \central A_p.$
The former groups are of exponent $p$, the latter ones are of exponent $p^2$.

It follows from the above that any extraspecial group of order $p^{2k+1}$ can be generated
by $2k+1$ elements $x_1, y_1, \ldots , x_k, y_k$ and $z$. Any element of the group has a unique representation of the form $x_1^{i_1}y_1^{i'_1}\cdots x_k^{i_k}y_k^{i'_k}z^\ell$,
where $i_1,i'_1,\ldots,i_k,i'_k,\ell\in \Z_p$. Also, 
$G'=Z(G)=\{z^\ell|\ell\in\Z_p\}$.

\section{Reduction lemmas}\label{reductions}
%We define the {\em subgroup problem} which is an extension of the $\HSP$.
%Let $G$ be a finite group 
%given by generators and relations,
%and let $f$ be a function, accessible via an oracle, which
%maps $G$ into a finite set. Let $H$ be a subgroup of $G$ defined via the function $f$. The problem
%$\SP(G, f, H)$ is the problem of finding 
%generators for 
%$H$. 
%In general $f$ doesn't have to be a hiding function for a subgroup $H$ of $G$, but clearly,
%in that case $\HSP(G, f)$ is just $\SP(G, f, H).$

%For $i=1,2$ let $G_i$ be a finite group, $f_i$ a function mapping $G_i$ into some finite set, and
%$H_i$ subgroup of $G_i$. We say that $\SP(G_1, f_1, H_1)$ is {\em reducible} to 
%$\SP(G_2, f_2, H_2)$ if the existence of a polynomial time quantum 
%procedure for solving $\SP(G_2, f_2, H_2)$
%implies the existence of a polynomial time quantum procedure for $\SP(G_1, f_1, H_1)$.
Our results leading to our main technical contribution can be the best described via
a series of reduction lemmas.
%reductions involving different instances of the $\SP$.
\begin{lemma}\label{lemma:firstreduction}
Let $G$ be an extraspecial $p$-group, and let us given an oracle $f$ which
hides the subgroup $H$ of $G$. Then
finding $H$ is efficiently reducible to find $HG'$.
\end{lemma}
\begin{proof}
Since $G'$ is a cyclic group of prime order, 
either $G' \subseteq H$ or $G' \cap H = \{1\}$. It is simple to decide which one of this cases
holds by checking if $f(z) = f(1)$. If $G' \subseteq H$ then $H = HG'$, and therefore the algorithm
which finds $HG'$ yields immediatly $H$.

If $G' \cap H = \{1\}$ then we claim that $HG'$ is abelian. To see this, it is sufficient to show that
$H$ is abelian, since $G'$ is the center of $G$.  Let 
$h_1$ and $h_2$ be two elements of $H$. Then there exists $\ell \in \Z_p$ such that 
$h_1 h_2= h_2 h_1 z^{\ell}$. This implies that $z^{\ell}$ is in $G' \cap H$ 
and therefore $z^{\ell} = 1$.

The restriction of the hiding function $f$ to the abelian subgroup $HG'$
of $G$ hides $H$.
Therefore the standard algorithm for solving the $\HSP$ in abelian groups 
applied to $HG'$ with oracle $f$ yields $H$.\qed
\end{proof}

We will show that finding $HG'$ can be efficiently reduced to the hidden subgroup problem in an abelian group.
%In a second step, we're going to reduce finding $HG'$ in $G$ to the construction of a function 
%$\overline{f}:G \rightarrow \C^{G^{4}}$ efficiently computable thanks to the original function $f$.
%Find $HG{'}$ in $G$ is difficult as $G$ is not abelian and so we can't apply the standard algorithm 
%directly. Let's see how to overcome this problem. 
For every element  $g = x_1^{i_1}y_1^{j_1} \ldots x_k^{i_k}y_k^{j_k}z^{\ell}$ of $G$, we denote 
by $\overline{g}$ 
%$\overline{x_1^{i_1}y_1^{j_1} \ldots x_k^{i_k}y_k^{j_k}z^{l}}$ for 
the element $x_1^{i_1}y_1^{j_1} \ldots x_k^{i_k}y_k^{j_k}$. 
%In the following, we'll  use $\cdot$ to refer to the law of G.
We define now the group $\overline{G}$ whose base set is $\{ \overline{g} : g \in G\}$. Observe that
this set of elements does not form a subgroup in $G$.
%$$\overline{G}=\{x_1^{i_1}y_1^{j_1} 
%\ldots x_k^{i_k}y_k^{j_k}, i_1 \ldots i_k \in \Z_p, j_1 \ldots j_k \in \Z_p\}$ 
To make $\overline{G}$ a group, its law is defined by 
$\overline{g_1} \ast \overline{g_2}=\overline{g_1  g_2}$ 
for all $\overline{g_1}$ and $\overline{g_2}$ in $\overline{G}$. 
It is easy to check that $\ast$ is well defined,
and is indeed a group multiplication. The group $\overline{G}$ is 
isomorphic to $G/G'$ and therefore is abelian. For our purposes a nice way to think about
$\overline{G}$ as a representation of $G/G'$ with unique encoding. In fact, it is also easy to check that
$\overline{G}$ is isomorphic to $\Z_p^{2k}$. Finally let us observe that 
$HG' \cap \overline{G}$ is a subgroup of $(\overline{G}, \ast)$ since
$HG'/G'$  is a subgroup of $G/G'$,

\begin{lemma}\label{lemma:secondreduction}
Let $G$ be an extraspecial $p$-group, and let us given an oracle $f$ which
hides the subgroup $H$ of $G$. Then
finding $HG'$ is efficiently reducible to find $HG' \cap \overline{G}$ in $\overline{G}$ .
%finding $H$ in $G$ is efficiently reducible to finding $HG'$ in $G$.
\end{lemma}
\begin{proof}
Since $HG'=(HG' \cap \overline{G}) G'$, a generator set 
of $HG'$ in $G$ is composed of a generator set of $HG' \cap \overline{G}$ in
$\overline{G}$ together with $z$.\qed
\end{proof}

%Since $HG'=(HG' \cap \overline{G})\cdot G'$, find $HG'$ in $G$ can be reduced to find 
%$HG' \cap \overline{G}$ in $(\overline{G}, \ast)$. Since $\overline{f}$ hides $HG'$, the restriction 
%of $\overline{f}$ to $\overline{G}$ hides $HG' \cap \overline{G}$ in $\overline{G}$ which is abelian, 
%and so we can find $HG'$.
The group $\overline{G}$ is abelian but we don't have a hiding function for 
$HG' \cap \overline{G}$. The main technical result of our paper is that using
the hiding function $f$ for $H$ in $G$, we will be able to implement an efficient
quantum 
%Gabor: \em
{\em hiding procedure} 
for $HG'$ in $G$.
Our last reduction lemma just states that this is sufficient for finding $HG' \cap \overline{G}$.

\begin{lemma}\label{lemma:thirdreduction}
Let $G$ be an extraspecial $p$-group, and let us given an oracle $f$ which
hides the subgroup $H$ of $G$. If we have an efficient quantum procedure (using $f$) 
which hides $HG'$ in $G$ then we can find efficiently $HG' \cap \overline{G}$ in $\overline{G}$ .
%finding $H$ in $G$ is efficiently reducible to finding $HG'$ in $G$.
\end{lemma}
\begin{proof}
The procedure which hides $HG'$ in $G$ hides also
$HG' \cap \overline{G}$ in $\overline{G}$. Since $\overline{G}$ is abelian,
Fact~\ref{fact:hsp} implies that we can find efficiently 
$HG{'} \cap \overline{G}$.\qed
\end{proof}
%We are now in the main part of the paper, the point where we have to 
%build $\overline{f}:G \rightarrow \C^{G^4}$ that hides $HG'$.
Our first theorem 
is the consequence of these  three lemmas. It says that
if in an extraspecial group we succeed to transform the oracle hiding  the subgroup $H$ into a 
quantum procedure hiding $HG'$ then we can determine $H$. This reduction
is the basis of our algorithm.
\begin{theorem}\label{theorem:threelemmas}
Let $G$ be an extraspecial $p$-group, and let us given an oracle $f$ which
hides the subgroup $H$ of $G$. If we have an efficient quantum procedure (using $f$) 
which hides $HG'$ in $G$ then $\HSP(G, f)$ can be solved efficiently. 
%finding $H$ in $G$ is efficiently reducible to finding $HG'$ in $G$.
\end{theorem}
Observe that if $G' \subseteq H$ then $HG' = H$, and therefore the following corollary is immediate.
\begin{corollary}\label{corollary:threelemmas}
Let $G$ be an extraspecial $p$-group, and let us given an oracle $f$ which
hides the subgroup $H$ of $G$. If  $G' \subseteq H$ 
then we can solve efficiently
$\HSP(G, f)$.
%finding $H$ in $G$ is efficiently reducible to finding $HG'$ in $G$.
\end{corollary}

%Observe that this result already implies that we can find efficiently those subgroups which
%contain the centrum of the group. Indeed, if $G' \subseteq H$ then $HG' = H$ and therefore
%$f$ itself is the hiding procedure for $HG'$.

\section{The algorithm}\label{algorithm}

We now describe the quantum algorithm which solves the $\HSP$ in extraspecial groups. 
In fact, 
we will deal separately with three cases: 
groups of constant exponent, 
groups of exponent $p$ when $p$ is large, and
groups of exponent $p^2$ when $p$ is large.
The  case of constant exponent is actually not new, it follows from a general
result in \cite{ims03}. Nevertheless, for the sake of completeness we show how a simplified version
of the algorithm for the second case works here.
The algorithm for extraspecial groups of exponent $p$ that goes to infinity
is our main result. Finally, the case of groups of exponent $p^2$ can be easily reduced to the
case of groups of exponent $p$. These results are summarized in our main theorem.

\begin{theorem}\label{theorem:main}
Let $G$ be an extraspecial $p$-group, and 
let us given an oracle $f$ which
hides the subgroup $H$ of $G$. Then there is an efficient quantum procedure which
finds $H$.
\end{theorem}

\subsection{Groups of constant exponent}

In Theorem 9 of \cite{ims03} it is proven that in general the $\HSP$ can be solved 
by a quantum algorithm in polynomial time in the size of the input and the cardinality of $G'$.
This includes the case of extraspecial groups of constant exponent. Nonetheless, for the
sake of completeness we describe here an efficient procedure, similar in spirit
to the one used for the next case but much simpler.

First remark that for every $a \in G$, the set $\{ \ket{aHG'  \cdot g} : g \in G \}$ is hiding for $HG'$ in $G$.
%This is proven as the analogous claim in Theorem~\ref{theorem:exponentp}.
%If $g_1$ and $g_2$ are from different cosets of $HG'$ then $\supp(\ket{aHG' g_1})$ and
%$\supp(\ket{aHG' g_2})$ are included in different cosets of $HG'$ and are disjoint.
%Therefore the states $\ket{aHG' g_1}$ and $\ket{aHG' g_2}$ are orthogonal.
%If $g_1$ and $g_2$ are in the same coset of $HG'$ then $g_1 = g g_2$ for some $g \in HG'$.
%Then $\ket{aHG' g} = \ket{aHG' }$, and therefore $\ket{aHG' g_1} = \ket{aHG' g_2}$.
The efficient hiding procedure for $HG'$
computes, for some $a \in G$, the superposition 
$ \frac{1}{\sqrt{p}} \sum _{u \in \Z_p} \ket{u} \ket{aHG'_u}$ 
which by Lemma~\ref{lemma:superposition} of Section~\ref{important}
can be done efficiently. Then the first register is measured.
This is repeated until the result of the observation is 0. Since $p$ is 
constant, after a constant number of iteration the superposition 
$\ket{0} \ket{aHG'_0} = \ket{0} \ket{aHG'}$ is created and finally  $\ket{aHG'  \cdot g}$
is computed.

Observe that this simplified approach can not work for large exponents 
since $p$, the expected
number of iterations, is not polynomial in the size of the input.

\subsection{Groups of exponent $p$ when $p$ is large}\label{important}

For every $u \in \Z_p$, let $\ket{G'_u}=
\frac{1}{\sqrt{p}} \sum_{i \in \Z_p} \omega^{-u i} \ket{z^i}$ 
and observe that $\ket{G'_u \cdot  z}=\omega^u \ket{G'_u}$.

\begin{lemma}\label{lemma:superposition}
There is an efficient quantum procedure which creates 
%the superposition
$ \frac{1}{\sqrt{p}} \sum _{u \in \Z_p} \ket{u} \ket{aHG'_u}$ where $a$ 
is a random element from $G$. 
\end{lemma}
\begin{proof}
We start with $\ket{0}\ket{0}\ket{0}$.
Since we have access to the hiding function $f$,
we can create the superposition $\frac{1}{\sqrt{|G|}}\sum_{g \in G} \ket{0} \ket{g} \ket{f(g)}$. 
Observing and discharging the third
register we get $\ket{0} \ket{aH}$ for a random element~$a$.
Applying the Fourier transform over $\Z_p$ to the
first register gives $\ket{\Z_p}\ket{aH}$.
Multiplying  the second register by $z^{-i}$ when $i$ is the content of the first one results in
$ \frac{1}{\sqrt{p}} \sum _{i \in \Z_p} \ket{-i} \ket{a H z^{i}}$. A final Fourier transform in the first register
creates the required superposition.\qed
\end{proof}

%Gabor: modified
%For $j=1, \ldots, p-1$, we define the endomorphisms $\phi_j$ over $G$ by 
%$\phi_j(x_i)=x_i^j$, $\phi_j(y_i)=y_i^j$ when $i \in \{1, \ldots, k\}$, and $\phi_j(z)=z^{j^2}$. The maps
%$\phi_j$ are in fact automorphisms of $G$ since the elements 
%$x_1^j, y_1^j, \ldots x_k^j, y_k^j, z^{j^2}$ generate the group $G$ and satisfy the defining relations.
For $j=1, \ldots, p-1$, we define the automorphisms $\phi_j$ of $G$ 
mapping $x_i$ to $x_i^j$, $y_i$ to $y_i^j$ and $z$ to $z^{j^2}$ when 
$i \in \{1, \ldots, k\}$. These maps (defined on generators) 
extend in fact to automorphisms of $G$ since the elements 
$x_1^j, y_1^j, \ldots x_k^j, y_k^j, z^{j^2}$ 
generate the group $G$ and satisfy the defining relations.

In our next lemma we claim that the states $\ket{aHG'_u}$ are eigenvectors of
the group action of multiplication from the right by $\phi_j(g)$, 
whenever $g$ is from $HG'$. Moreover, the 
corresponding eigenvalues are some powers of the 
root of the unity, the exponent does not depend on $a$, and the dependence on $u$ and $j$ 
is relatively simple.

\begin{lemma}\label{lemma:action}
We have
\begin{enumerate}
\item  $\forall h \in H, \exists \ell \in \Z_p, \forall a \in G, \forall u \in \Z_p, \forall j \in \Z_p^*, ~~
\ket{aHG'_u \cdot  \phi_j(h)}=\omega^{u(j-j^2)\ell} \ket{aHG'_u},$

\item $ \forall a \in G, \forall u \in \Z_p, \forall j \in \Z_p^*, ~~
\ket{aHG'_u \cdot \phi_j(z)}=\omega^{uj^2} \ket{aHG'_u}.$
\end{enumerate}
\end{lemma}

\begin{proof}
To begin with let's remark that for $h \in H$, we have $\ket{aHG'_u \cdot h}=\ket{aHG'_u}$ 
and that $\ket{a H G'_u \cdot z}= \omega^u \ket{a H G'_u}$.

To prove the first part, let $h$ be an element of $H$. Then $\phi_j(h)=h^jz^{t}$ where
$t$ depends on $h$ and $j$. We will show that $t = (j-j^2)\ell$
where $\ell$ depends only on $h$. This will imply the claim.

Let $j_0$ be a fixed primitive element of $\Z_p^*$. Then
$\phi_{j_0}(h)=h^{j_0}  z^s$, for some $ s \in \Z_p$.
We set $\ell=s(j_0-j_0^2)^{-1}$, and $k=h  z^{\ell}$. 
Then $\phi_{j_0}(k)=h^{j_0}  z^{\ell(j_0-j_0^2)} z^{\ell j_0^2}=k^{j_0}$.
Therefore $\phi_{j}(k) = k^j$ and
$\phi_j(h)=\phi_j(k) \phi_j(z^{-\ell})=h^jz^{\ell(j-j^2)}$. The proof of the second part is immediate.\qed
%Since $\phi_j(h)=h^jz^{l(j-j^2)}$we have $\ket{aHG'_u \cdot \phi_j(h)}=\ket{aHG'_u \cdot h^j 
%\cdot z^{l(j-j^2)}}=\omega^{u(j-j^2)l} \ket{aHG'_u}$, which proves the first part of the claim.
%
%The proof of the second part is easy:
%
%$\ket{aHG'_u \cdot \phi_j(z)}=\ket{aHG'_u \cdot z^{j^2}}=\omega^{uj^2} \ket{aHG'_u}$.
\end{proof}

The principal idea now is to
take several copies of the states $\ket{a_iHG'_{u_i}}$ and choose 
$j_i$ so that the product 
of the corresponding 
%Gabor:
eigenvalues
%eigenvectors 
becomes
the unity. Therefore the actions
$\phi_j(g)$, when $g$ is from $HG'$, will not modify the combined state. It turns out  
that we can achieve this with four copies.

For $\overline{a}=(a_1,a_2,a_3,a_4) \in G^4$, 
$\overline{u}=(u_1,u_2,u_3,u_4) \in \Z_p^4$, 
$\overline{j}=(j_1,j_2,j_3,j_4) \in ({\Z_p^*})^4$ and $g \in G$, we define the quantum state
$\ket{ \Psi_g^{\overline{a}, \overline{u}, \overline{j}}}$ in $\C^{G^4}$ by
$$\ket{ \Psi_g^{\overline{a}, \overline{u}, \overline{j}}}  =
\ket{a_1HG'_{u_1} \cdot \phi_{j_1}(g),a_2HG'_{u_2} \cdot \phi_{j_2}(g),a_3HG'_{u_3} \cdot \phi_{j_3}(g),a_4HG'_{u_4} \cdot \phi_{j_4}(g)}.$$

Our purpose is to find an efficient procedure to generate triples 
$(\overline{a}, \overline{u}, \overline{j})$ such that 
for 
every $g \in HG'$ we have 
$\ket{ \Psi_g^{\overline{a}, \overline{u}, \overline{j}}}
=
\ket{a_1HG'_{u_1},a_2HG'_{u_2},a_3HG'_{u_3},a_4HG'_{u_4}} $. We call such triples
{\em appropriate}. The reason to look for appropriate triples is that they lead to hiding
sets for $HG'$ in $G$ as stated in the next lemma.

\begin{lemma}\label{lemma:appropriate}
If $(\overline{a}, \overline{u}, \overline{j})$ is an appropriate triple then 
$\{ \ket{ \Psi_g^{\overline{a}, \overline{u}, \overline{j}}} : g \in G \}$ is hiding for $HG'$ in $G$.
\end{lemma}

\begin{proof}
To see this, first observe that $HG'$ is a normal subgroup of $G$. 
If $g_1$ and $g_2$ are in different cosets of $HG'$ in $G$ then
for every $j \in \Z_p^*$,
the elements $\phi_j(g_1)$ and $\phi_j(g_2)$ are in different cosets of 
$HG'$ in $G$ since $\phi_j$ is an automorphism of $G$. Also, for every
$a \in G$ and for every $u \in \Z_p$ we have 
$\supp(\ket{aHG'_u}) =  \supp (  \ket{aHG'})$, and therefore
$\supp(\ket{aHG'_u \cdot \phi_j(b)})$ and 
$\supp(\ket{aHG'_u \cdot \phi_j(b^{'})})$ are included in different cosets and are disjoint. 
Thus for every $\overline{a} \in G^4, 
\overline{u} \in \Z_p^4$ and $ \overline{j}  \in ({\Z_p^*})^4$,
the states $\ket{ \Psi_{g_1}^{\overline{a}, \overline{u}, \overline{j}}}$ and 
$\ket{ \Psi_{g_2}^{\overline{a}, \overline{u}, \overline{j}}}$ are orthogonal.

If $g_1$ and $g_2$ are in the same coset of $HG'$ then $g_1 = g g_2$ for some $g \in HG'$,
and $\phi_{j_i}(g_1) = \phi_{j_i}(g) \phi_{j_i}(g_2)$. Thus
$\ket{ \Psi_{g_1}^{\overline{a}, \overline{u}, \overline{j}}} =
\ket{ \Psi_{g g_2}^{\overline{a}, \overline{u}, \overline{j}}}  =
\ket{ \Psi_{g_2}^{\overline{a}, \overline{u}, \overline{j}}}$.\qed
\end{proof}

Let us now address the question of existence of appropriate triples and efficient ways to
generate them. Let $(\overline{a}, \overline{u}, \overline{j})$ be an arbitrary element of 
$G^4 \times \Z_p^4 \times (\Z_p^*)^4$, and
let $g$ be an element of  $HG'$. Then $g=h z^t$ for some $h \in H$ and 
$t \in \Z_p$, and $\phi_{j_i}(g)=\phi_{j_i}(h) \phi_{j_i}(z^t)$ for $i=1, \ldots, 4$. 
By Lemma~\ref{lemma:action} there exists $\ell$ such 
that $\ket{a_iHG'_{u_i} \cdot \phi_j(h)}= 
\omega^{u_i(j_i-j_i^2)\ell}\ket{a_iHG'_{u_i}}$ and 
$\ket{a_iHG'_{u_i} \cdot \phi_j(z^t)}= \omega^{u_i j_i^2 t}\ket{a_iHG'_{u_i}},$ and therefore
$$
\ket{ \Psi_g^{\overline{a}, \overline{u}, \overline{j}}}
= \omega^{\sum_{i=1}^4 ( u_i(j_i-j_i^2)\ell +  u_i j_i^2 t ) }
\ket{a_1HG'_{u_1},a_2HG'_{u_2},a_3HG'_{u_3},a_4HG'_{u_4}}.
$$
We say that $\overline{u} \in \Z_p^4$ is {\it good} if the following system of
quadratic equations has a nonzero solution: 
\begin{equation}
\begin{cases}
\sum_{i=1}^4  u_i(j_i-j_i^2) & =~~ 0 \\
\sum_{i=1}^4  u_ij_i^2  & =~~ 0,
\end{cases}
\end{equation}
and we call a solution $\overline{j}$ a {\em witness} of $u$ being good. It should be clear that
for every $\overline{u}$, if $\overline{u}$ is good and $\overline{j}$ witnesses that then
$(\overline{a}, \overline{u}, \overline{j})$ is an appropriate triple.

The next lemma states that a random $\overline{u}$ is good with constant probability, and that
in this case one can find efficiently $\overline{j}$ witnessing that.

\begin{lemma}\label{lemma:probability}
For every $\overline{a} \in G^4$, we have
$$\prob{\overline{u} \in \Z_p^4}  { 
\overline{u} \mbox{~{\rm is good}}
 }\geq 
(p - 9)/2p.$$
Moreover, when $\overline{u}$ is good a witness $\overline{j}$ can be found efficiently.
\end{lemma}

\begin{proof}

Let us simplify system (1) to the equivalent system

\begin{equation}
\begin{cases}
\sum_{i=1}^4  u_ij_i^2 & =~~ 0 \\
\sum_{i=1}^4  u_ij_i  & =~~ 0.
\end{cases}
\end{equation}
To solve (2),
we take $j_3=1$ and $j_4=-1$, and we set $v=u_3+u_4$ and $w=u_3-u_4$. We will show that
for random $(u_1,u_2,v,w)  \in \Z_p^4$, the reduced system (3) 
has a solution $(j_1,j_2) \in (\Z_p^*)^2$
with probability at least $(p - 9)/2p$, and that 
the solution is easy to find:

%\begin{tabular} {ccc}
% $u_1j_1^2+u_2j_2^2$ & $=$ & $-v$\\
 %$u_1j_1+u_2j_2$ & $=$ & $-w$\\
 %\end{tabular}
 
\begin{equation}
\begin{cases}
u_1j_1^2+u_2j_2^2 & =~~ -v \\
u_1j_1+u_2j_2 & =~~ -w.
\end{cases}
\end{equation}

With probability at least $1 - 3p$ we have $u_1 \neq 0$, $u_2 \neq 0$, $u_1+u_2 \neq 0$.
In that case we can substitute $j_2=-\frac{w+u_1j_1}{u_2}$ in the first equation and get
in $j_1$ the quadratic equation  $(u_1u_2 + u_1^2)j_1^2 +2u_1w j_1 +(w^2+ vu_2)=0$.
It is a non degenerate quadratic equation whose discriminant
$D=-4u_1u_2(w^2+(u_2+u_1)v)$ is uniformly distributed in $\Z_p$ since it is linear in $v$.
Therefore $D$ is a quadratic residue with probability $(p-1)/2p$,
%Gabor:
 and we can efficiently compute a square root of $D$ modulo $p$
(see, for example,  subsection 13.3.1 of \cite{shoup05}). 
We also have to ensure that 
$j_2 \neq 0$.
If $j_2$ is zero, then $w^2=-vu_1$, which happens with probability $1/p$.
Therefore the probability of finding a solution $(j_1,j_2) \in (\Z_p^*)^2$ is at least
$(p-1)/2p - 4/p$.\qed
% begin suppress
\suppress{
For $u_2 \neq 0$ let's $j_2=-\frac{w+u_1j_1}{u_2}$, we need to solve the quadratic equation in $j_1$, $(u_1u_2 + u_1^2)j_1^2 +2u_1w j_1 +(w^2+ vu_2)=0$ of discriminant $D=-4u_1u_2(w^2+(u_2+u_1)v)$. This equation has a non zero solution at least when $D$ is a non zero square. If $u_1 \neq 0$, $u_2 \neq 0$, $u_1+u_2 \neq 0$ then the equation is a non degenerate quadratic one with uniformly random discriminant as it is linear in $v$. That happens with probability at least $\frac{p-1}{2p}-\frac{3}{p}$. In those solution we're going to keep only the ones for which $j_2$ is non zero. If $j_2$ is zero, then $w^2=-vu_1$, what happens with probability $\frac{1}{p}$ when $u_1$ is non zero. The probability to find a good solution to our system of equation is then at least $\frac{1}{2}-\frac{9}{2p}$.
}
% end suppress
\end{proof}

\begin{theorem}\label{theorem:exponentp}
Let $G$ be an extraspecial $p$-group of exponent $p$, where $p$ grows with the input size, and 
let us given an oracle $f$ which
hides the subgroup $H$ of $G$. Then there is an efficient quantum procedure which
hides $HG'$ in $G$.
\end{theorem}
\begin{proof}
We describe the efficient hiding procedure.
It computes, for some $\overline{a} \in G^4$, the superposition 
$$ \frac{1}{{p}^2} \bigotimes_{i=1}^4   \sum _{u_i \in \Z_p} \ket{u_i} \ket{a_iHG'_{u_i}},$$
which by Lemma~\ref{lemma:superposition} can be done efficiently, and then 
it measures the registers for the $u_i$. This is repeated until a good 
$\overline{u} \in \Z_p^4$ is measured. By Lemma~\ref{lemma:probability}, this requires a
constant expected number of iterations. Also, when a good $\overline{u}$ is measured, it
finds efficiently a solution $ \overline{j}  \in ({\Z_p^*})^4$ for system (1).
Such a triple $(\overline{a}, \overline{u}, \overline{j})$ is appropriate, and therefore by
Lemma~\ref{lemma:appropriate}
$\{ \ket{ \Psi_g^{\overline{a}, \overline{u}, \overline{j}}} : g \in G \}$ is hiding for $HG'$ in $G$. 
Using the
additional input $\ket{g}$, the procedure finally computes 
$ \ket{ \Psi_g^{\overline{a}, \overline{u}, \overline{j}}} $.\qed
\end{proof}

The proof of Theorem~\ref{theorem:main} in that case follows from Theorem~\ref{theorem:threelemmas} and Theorem~\ref{theorem:exponentp}.

\subsection{Groups of exponent $p^2$ when $p$ is large}
Here we deal with the group
$G = A_p \central  (\multicentral{i=1}{k-1}{H_p}),$ where we start with a function $f$ hiding 
some subgroup $H$.
As in Lemma~\ref{lemma:firstreduction}, we will distinguish the cases when $G' \subseteq H$ 
and when $G' \cap H = \{e\}$. The first case is already taken care of by 
Corollary~\ref{corollary:threelemmas}.

If $G' \cap H = \{e\}$ then $H$ contains only elements whose order is at most $p$.
Indeed an element of order $p^2$ cannot be in $H$ since the $p^{\mbox{th}}$
power of such an
element is in $G'$. Therefore $H$ is a subgroup of $K = \langle  
y_1, x_2, y_2,  \ldots , x_{k}, y_k, z \rangle$,
where $x_1$ is the unique generator of order $p^2$ of $G$.
The subgroup $K$ is also (isomorphic to) a subgroup of
$\multicentral{i=1}{k}{H_p}$. We claim that we can extend the restriction of $f$ to $K$ into 
a function $F$ defined on
the whole group
$\multicentral{i=1}{k}{H_p}$  that also  hides $H$. Such an 
extension can be defined for example as 
$F(x_1^{i_1} y_1^{j_1} \ldots x_k^{i_k} y_k^{j_k} z^{\ell})=
(i_1,f(y_1^{j_1} \ldots x_k^{i_k} y_k^{j_k} z^{\ell}))$, and it is easy to see that it is indeed a hiding
function. Therefore the problem is reduced to the HSP in extraspecial groups of exponent $p$.

\section{Concluding remarks}
The main technical contribution of the present paper
is a quantum procedure which hides $HG'$
in an extrapsecial $p$-group $G$ where $p$ is a large
prime. We remark
that it is possible to present the proof of its correctness in terms
of irreducible representations of $G$. However, 
the present approach 
is shorter and it does not make use of concepts of  
noncommutative representation theory. Finally, our method can in turn be extended to
finding hidden subgroups efficiently in arbitrary finite
two-step nilpotent groups, that is groups
$G$ satisfying $G'\leq Z(G)$. This extension will
be the subject of a subsequent paper.

\subsubsection*{Acknowledgment.}

The authors are grateful to P\'eter P\'al P\'alfy for 
his useful remarks and suggestions.


\begin{thebibliography}{20}

\bibitem{asc00}
M.~Aschbacher.
\newblock {\em Finite Group Theory}.
\newblock Cambridge University Press, 2000.


\bibitem{bcv05}
D.~Bacon, A.~Childs, and W.~van~Dam.
\newblock From optimal measurement to efficient quantum algorithms 
for the hidden subgroup problem over semidirect product groups.
\newblock In {\em Proc. 46th IEEE FOCS}, pages 469--478, 2005.

%\bibitem{bea97}
%R.~Beals.
%\newblock Quantum computation of {Fourier} transforms over symmetric groups.
%\newblock In {\em Proc. 29th ACM STOC}, pages 48--53, 1997.

%\bibitem{bs84}
%L.~Babai and E.~Szemer\'edi.
%\newblock On the complexity of matrix group problems {I}.
%\newblock In {\em Proc. 25th FOCS}, pages 229--240, 1984.

\bibitem{crss97}
A.~Calderbank, E.~Rains, P.~Shor and N.~Sloane.
\newblock Quantum error correction and orthogonal geometry.
\newblock {\em Phys. Rev. Lett.}, 78:405--408, 1997. 
%quantph/9608006

\bibitem{crss98}
A.~Calderbank, E.~Rains, P.~Shor and N.~Sloane.
\newblock Quantum error correction via codes over GF(4).
\newblock {\em IEEE Transactions on Information Theory}, 44(4):1369--1387, 1998. 

%\bibitem{cm01}
%K.~Cheung and M.~Mosca.
%\newblock Decomposing finite abelian groups.
%\newblock {\em J. Quantum Inf. Comp.}, 1(3), 2001.

%\bibitem{dhi03}
%W.~van Dam, S. Hallgren, and L.~Ip.
%\newblock Quantum algorithms for some hidden shift problems.
%\newblock In {\em Proc. 14th ACM-SIAM SODA}, 2003.

%\bibitem{eh00}
%M.~Ettinger and P.~H{\o}yer.
%\newblock On quantum algorithms for noncommutative hidden subgroups.
%\newblock {\em Adv. in Appl. Math.}, 25(3):239--251, 2000.

\bibitem{fimss03}
K. Friedl, G. Ivanyos, F. Magniez , M. Santha and P. Sen.
\newblock Hidden translation and orbit coset in quantum computing.
\newblock In {\em Proc. 35th ACM STOC}, pages 1--9, 2003.

\bibitem{got97}
D.~Gottesman.
\newblock {\em Stabilizer Codes and Quantum Error Correction}.
\newblock PhD Thesis, Caltech, 1997.

\bibitem{gsvv01}
M.~Grigni, L.~Schulman, M.~Vazirani, and U.~Vazirani.
\newblock Quantum mechanical algorithms for the nonabelian 
          {Hidden Subgroup Problem}.
\newblock In {\em Proc. 33rd ACM STOC}, pages 68--74, 2001.

%\bibitem{hmr06}
%S.~Hallgren, C.~Moore, M.~R\" otteler, A.~Russell and P.~Sen.
%\newblock Limitations of quantum coset states for Graph Isomorphism.
%\newblock In {\em Proc. 38th ACM STOC}, pages 604--617, 2006.

\bibitem{hrt03}
S.~Hallgren, A.~Russell, and A.~{Ta-Shma}.
\newblock Normal subgroup reconstruction and quantum computation 
          using group representations.
%\newblock In {\em Proc. 32nd ACM STOC}, pages 627--635, 2000.
\newblock {\em SIAM J. Comp.}, 32(4):916--934, 2003.

\bibitem{hup83}
B.~Huppert.
\newblock {\em Endliche Gruppen}. Vol. 1,
\newblock Springer Verlag, 1983.

\bibitem{ims03}
G.~Ivanyos, F.~Magniez, and M.~Santha.
\newblock Efficient quantum algorithms for some instances 
          of the non-{A}belian hidden subgroup problem.
%\newblock In {\em Proc. 13th ACM SPAA}, pages 263--270, 2001.
\newblock {\it Int. J. of Foundations of Computer Science}, 14(5):723--739, 2003.

\bibitem{kit95}
A.~Kitaev.
\newblock Quantum measurements and the {A}belian {S}tabilizer {P}roblem.
\newblock Technical report, Quantum Physics e-Print archive, 1995.
\newblock {\tt http://xxx.lanl.gov/abs/quant-ph/9511026}.

\bibitem{ks06}
A.~Klappenecker, P.~K.~Sarvepalli.
\newblock Clifford Code Constructions of Operator Quantum Error Correcting Codes
\newblock Technical report, Quantum Physics e-Print archive, 2006.
\newblock {\tt http://xxx.lanl.gov/abs/quant-ph/0604161}.


%\bibitem{ksv02}
%A.~Kitaev, A.~Shen, and M.~Vyalyi.
%\newblock Classical and quantum computation.
%\newblock In {\em Graduate Studies in Mathematics}, volume~47. AMS, 2002.

%\bibitem{kup03}
%G.~Kuperberg,
%\newblock A subexponential-time quantum algorithm for the dihedral hidden subgroup.
%\newblock Technical report, Quantum Physics e-Print archive, 2003.
%\newblock {\tt http://xxx.lanl.gov/ abs/quant-ph/0302112}.
%\newblock {\em SIAM J. Comp.}, 35(1):170--188, 2005.

\bibitem{nc00}
M.~Nielsen and I.~Chuang.
\newblock {\em Quantum Computation and Quantum Information}.
\newblock Cambridge University Press, 2000.

\bibitem{mrrs04}
C. Moore, D. Rockmore, A. Russell, and L. Schulman.
\newblock The power of basis selection in Fourier sampling: 
Hidden subgroup problems in affine groups.
\newblock In {\em Proc. 15th ACM-SIAM SODA}, pages 1106--1115, 2004.

%\bibitem{mr05}
%C.~Moore, A.~Russell.
%\newblock The symmetric group defies strong Fourier sampling: Part II.
%\newblock ArXiv preprint quant-ph/0501066, 2005.

%\bibitem{mrs05}
%C.~Moore, A.~Russell, and L.~Schulman.
%\newblock The symmetric group defies strong Fourier sampling: Part I.
%\newblock ArXiv preprint quant-ph/0501056, 2005.

\bibitem{mos99}
M.~Mosca.
\newblock {\em Quantum Computer Algorithms}.
\newblock PhD Thesis, University of Oxford, 1999.

\bibitem{prb99}
M.~P\" uschel, M.~R\" otteler, and T.~Beth.
\newblock Fast quantum {F}ourier transforms for a class 
          of non-{A}belian groups.
\newblock In {\em Proc. 13th AAECC}, volume 1719, 
          pages 148--159. LNCS, 1999.

%\bibitem{pre98}
%J.~Preskill.
%\newblock Quantum information and computation.
%\newblock {\tt http://www.theory.caltech.edu/people/preskill/ ph229},
%          1998.

\bibitem{rrs05}
J.~Radhakrishnan, M.~R\" otteler and P.~Sen.
\newblock On the power of random bases in Fourier sampling: 
          hidden subgroup problem in the Heisenberg group.
\newblock In {\em Proc. 32nd ICALP}, LNCS vol. 3580, pages 1399--1411, 2005.

\bibitem{rb98}
M.~R\" otteler and T.~Beth.
\newblock Polynomial-time solution to the {Hidden Subgroup Problem} 
          for a class of non-abelian groups.
\newblock Technical report, Quantum Physics e-Print archive, 1998.
\newblock {\tt http://xxx.lanl.gov/abs/quant-ph/9812070}.

%\bibitem{rp00}
%{E. G.} Rieffel and W.~Polak.
%\newblock An introduction to quantum computing for non-physicists.
%\newblock {\em ACM Computing Surveys}, 32(3):300--335, 2000.

\bibitem{sho97}
P.~Shor.
\newblock Algorithms for quantum computation: {Discrete} logarithm and
          factoring.
\newblock {\em SIAM J. Comp.}, 26(5):1484--1509, 1997.

\bibitem{shoup05}
V.~Shoup.
\newblock {\em A Computational Introduction to Number Theory and Algebra.}
\newblock Cambridge University Press, 2005.


\bibitem{sim97}
D.~Simon.
\newblock On the power of quantum computation.
\newblock {\em SIAM J. Comp.}, 26(5):1474--1483, 1997.

%\bibitem{wat01}
%J.~Watrous.
%\newblock Quantum algorithms for solvable groups.
%\newblock In {\em Proc. 33rd ACM STOC}, pages 60--67, 2001.


\end{thebibliography}
\end{document}